\documentclass[%
 reprint,
 amsmath,amssymb,
 aps,
]{revtex4-2}

\usepackage{graphicx}% Include figure files
\usepackage{dcolumn}% Align table columns on decimal point
\usepackage{bm}% bold math
\usepackage{color}

\usepackage[caption=false]{subfig}

\renewcommand{\vec}[1]{\mathbf{#1}}

\begin{document}

\preprint{APS/123-QED}

\title{ Dynamic polarization and plasmons in kekul\'e-patterned graphene: \\ Signatures of the broken valley degeneracy }% Force line breaks with \\

\author{Sa\'ul A. Herrera}

\author{Gerardo G. Naumis}%
 \email{naumis@fisica.unam.mx}
\affiliation{%
Depto. de Sistemas Complejos, Instituto de F\'isica, \\ Universidad Nacional Aut\'onoma de M\'exico (UNAM)\\
Apdo. Postal 20-364, 01000, CDMX, M\'exico.
}%

\date{\today}% It is always \today, today,

\begin{abstract}

 The dynamic polarization for kekul\'e-patterned graphene is studied within the Random Phase Approximation (RPA).
It is shown how the breaking of the valley degeneracy by the lattice modulation is manifested through the dielectric spectrum,  the plasmonic dispersion, the static screening and the optical conductivity. The valley-dependent splitting of the Fermi velocities due to the kekul\'e distortion leads to a similar splitting in the dielectric spectrum of graphene, introducing new characteristic frequencies, which are given in terms of the valley-coupling amplitude. The valley coupling also splits the plasmonic dispersion, introducing a second branch in the Landau damping region.   Finally,  the signatures of the broken valley degeneracy in the optical conductivity are studied. The characteristic step-like spectrum of graphene is split into two half steps due to the onset of absorption in each valley occurring at different characteristic frequencies. Also, it was found an absorption phenomenon where a resonance peak related to intervalley transport emerges at a ``beat frequency'', determined by the difference between the characteristic frequencies of each valley.  Some of these mechanisms are expected to be present in other space-modulated 2D materials and suggest how optical or electrical response measurements can be suitable to detect spatial modulation.
\begin{description}
\item[Key words]
Kekul\'e, graphene, dielectric function
\end{description}
\end{abstract}

\maketitle

%\tableofcontents

\section{\label{Calculation}Introduction}
Space-modulated two dimensional materials (2D), are very interesting platforms  for novel physical phenomena \cite{MOGERA2020470,Wu_2020,NaumisReview,Chen2019,Yankowitz2018,Ni2015,Taboada2017,Ohta2012,Bistritzer,Zheng2016}. One of the most interesting systems is  kekul\'e-distorted graphene  \cite{Gamayun,Elias_2019,Wu2020,Tijerina_2019,Hoi_2019,Penglin_2019}, which has recently been observed in graphene sheets epitaxially grown over copper substrates \cite{Gutierrez2016}.
Tight binding models of Kekul\'e-Y (or Kek-Y) distorted graphene indicate a coupling of the charge carriers' pseudospin and orbital degrees of freedom \cite{Gamayun}. This results in the breaking of the valley degeneracy of graphene and two emerging species of massless Dirac fermions  \cite{Gamayun,Elias_2019}. Each species has a  different Fermi velocity, resulting in two Dirac cones with different slopes \cite{Gamayun}. A remarkable particularity about the Kek-Y phase in graphene is that both cones share the same Dirac point, as graphene's Brillouin zone is folded due to the increased size of the unitary cell. In graphene, the two nonequivalent Dirac cones are far away in momentum space, implying that intervalley transport between cones is forbidden at low energies. This is no longer the case in the  Kek-Y distorted phase, which makes it possible to access the valley degree of freedom in graphene. In fact, Kek-Y distorted graphene has been proven to be a potential platform to obtain strain-controlled valley-tronics, as the distance between valleys can be tuned externally \cite{Elias_2019}. As an example, a ballistic graphene-based valley field-effect transistor has been recently proposed \cite{Wang2020}.
More recently, it has been reported that the enabling of low-energy intervalley transport due to the Kek-Y distortion introduces an absorption peak in the optical gap of graphene \cite{Naumis2020}. This peak can be tuned in frequency and amplitude by changing the carrier density, making this phase a potential candidate for graphene-based optical modulators \cite{OpticalModulators,Liu2011,Andersen:10,Lee:12,Li2014,Luo2015,Liu_2013}, which rely on the highly tunable optical properties of graphene. 
From a topological point of view, kekul\'e patterned graphene can be considered as an extension of the Su-Schrieffer-Heeger model \cite{Fulde,TimeSSHModel}.  Mechanical strain on patterned graphene based heterostructures also leads to interesting topological effects \cite{Lazlo_2020}.
Also, the kekul\'e distortion has been proposed as a possible mechanism behind superconductivity in magic-angle twisted bilayer graphene \cite{Bitan_2010,Hoi_2019}, and  multiflavor Dirac fermions were predicted to emerge in kekul\'e graphene bilayers \cite{Tijerina_2019}. 
Moreover, it is possible to produce such pattern in other kinds of non-atomic systems, as with mechanical waves  in solids \cite{Mendez2020}, or in acoustical lattices, where topological Majorana modes were observed \cite{Penglin_2019}.  Additionally, kekul\'e ordering can be produced in photonic \cite{Photonic_2019}, polaronic \cite{Cerda2013} and atomic systems \cite{Atomic_2019}.
Therefore, the kekul\'e bond order is among one of the most interesting phases resulting from strain in a 2D material
\cite{NaumisReview}, having substantial potential for a wide range of applications \cite{Gamayun,Elias_2019,Naumis2020,Wu2020}.

The aim of this work is to study the consequences of the kekul\'e distortion in the dielectric response, static screening, optical conductivity and plasmonics of graphene and to gain insight into how the spatial modulation in similar 2D materials can be detected and characterized through optical and electrical measurements. We use as a starting point the tight binding model reported in Ref. \cite{Gamayun}. We focus on the Kek-Y phase, in which the Dirac cones $K$, $K'$ fold on top of each other and preserve the gapless dispersion. The layout of this work is the following. First we introduce the model in Sec. \ref{Sec.Model} and we make a general analysis of the polarizability in Sec. \ref{Sec.Polarizability}. In Secs. \ref{Sec.Plasmons} and \ref{Sec.Screening}, we study the effect of the kekul\'e distortion on the plasmon dispersion and the static screening, respectively, then a brief discussion on the conductivity is given in Sec. \ref{Sec.Conductivity}. Lastly, we give some general conclusions.

\section{Hamiltonian model for Kek-Y distorted graphene}\label{Sec.Model}

\begin{figure}[t]
\includegraphics[width=0.45\textwidth]{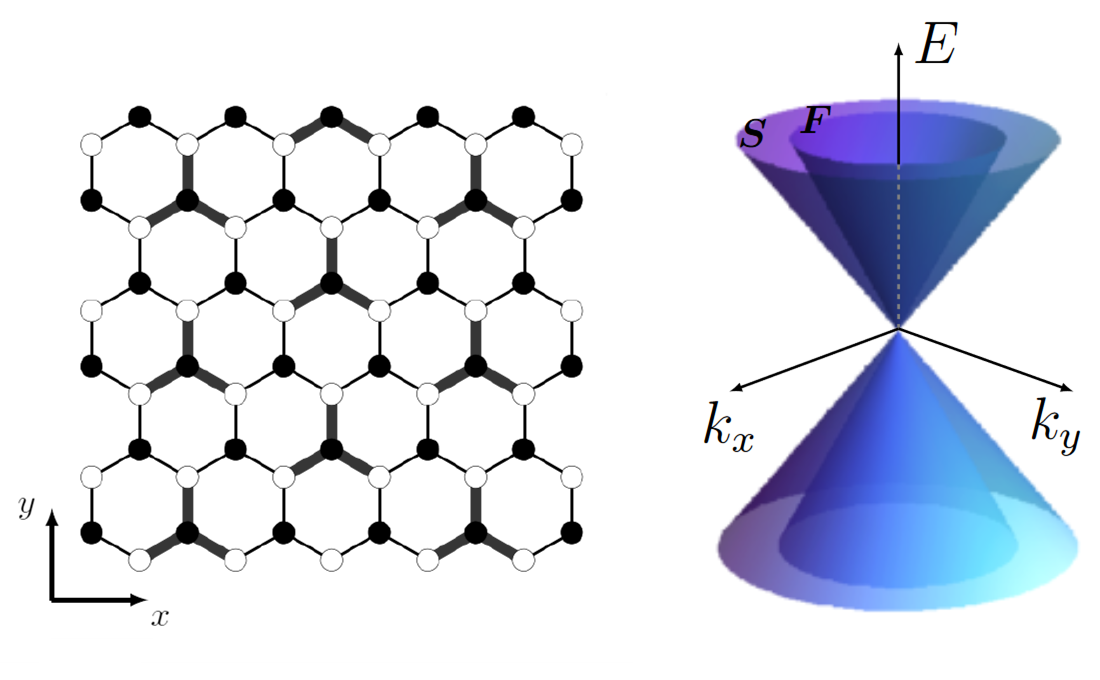}
\caption{\label{Fig:DispersionCone}To the left: graphene lattice with a type-Y kekul\'e modulation. To the right: energy dispersion of low-energy electronic excitations around the Dirac point in Kek-Y modulated graphene. The letter $S$ labels the slow cone, which has a slope given by the velocity $v_0(1-\Delta_0)$. The letter $F$ labels the fast cone, associated with the velocity $v_0(1+\Delta_0)$.}
\end{figure}

The graphene lattice with Kek-Y modulation is depicted in Fig. \ref{Fig:DispersionCone}, where the Y-shaped alternation of strong and weak bonds is shown. According to Gamayun et. al. \cite{Gamayun}, the low-energy  Hamiltonian for Kek-Y distorted graphene is given by the following $4\times4$ matrix,

\begin{equation}\label{Eq:HamiltonianMatrix}
    H =  \left( \begin{array}{ccc}
v_0\boldsymbol{p}\cdot\boldsymbol{\sigma} & \tilde{\Delta} Q_\chi\\
\tilde{\Delta}^\ast Q_\chi^\dagger & v_0 \boldsymbol{p}\cdot\boldsymbol{\sigma}  \end{array} \right),
\end{equation}
where $\tilde{\Delta}$ is the energy coupling amplitude due to the bond-density wave which describes the kekul\'e textures  and $\boldsymbol{\sigma}=(\sigma_x,\sigma_y)$ is a set of Pauli matrices. The Kek-Y texture coupling between Dirac Hamiltonians is given by the operator $Q_\chi=v_0(\chi p_x -i p_y)\sigma_0$ with
$|\chi|=1$ and $\sigma_0$ the identity. To avoid extra phases and for simplicity we consider a real $\tilde{\Delta}=\Delta_0$ and $\chi=1$, as  a complex $\tilde{\Delta}$ and $\chi=-1$ are equivalent upon an unitary transformation \cite{Gamayun}. In what follows we will also take $\hbar=1$. These considerations lead to the Hamiltonian,

\begin{equation}\label{Eq:HamiltonianMatrix2}
    H =  v_0 \left( \begin{array}{ccc}
\boldsymbol{k}\cdot\boldsymbol{\sigma} & \Delta_0 (k_x-i k_y)\sigma_0 \\
 \Delta_0 (k_x+i k_y)\sigma_0 &  \boldsymbol{k}\cdot\boldsymbol{\sigma}  \end{array} \right),
\end{equation}

or $H=v_0(\boldsymbol{k}\cdot\boldsymbol{\sigma})\otimes \tau_0 + v_0\Delta_0 \sigma_0 \otimes (\boldsymbol{k}\cdot\boldsymbol{\tau})$, with $\boldsymbol{\tau}=(\tau_x,\tau_y)$ defining a second pair of Pauli matrices, $\tau_0$ the unitary matrix, and $v_0$ the Fermi velocity of pristine graphene.

The spectrum  resulting from such Hamiltonian is given by,
\begin{equation}\label{Eq:Dispersion}
\epsilon_{\vec{k}\alpha}^\beta=\alpha s_\beta v_0  k
\end{equation}
where $\alpha= 1$ labels the conduction band and $\alpha= -1$ labels the valence band. The label $\beta=\pm 1$ is used to define  two velocities, $s_\beta=(1+\beta\Delta_0)$, and $k=\sqrt{k_x^2 + k_y^2}$. 
Therefore, as shown in Fig. \ref{Fig:DispersionCone}, the energy dispersion of the kekul\'e pattern folds graphene's $K$ and $K'$ valleys into the $\Gamma$ point of the superlattice Brillouin zone. This results in a ``fast" cone with Fermi velocity $v_0(1+\Delta_0)$, corresponding to $\beta=1$, and a ``slow" cone with Fermi velocity $v_0(1-\Delta_0)$, corresponding to $\beta=-1$. We label these cones as $F$ and $S$, respectively, leading to the the $F$ cone being described by the dispersion $\epsilon_{\vec{k},\pm }^{+}=\pm v_0(1+\Delta_0)k$ and the $S$ cone  by $\epsilon_{\vec{k},\pm }^{-}=\pm v_0(1-\Delta_0)k$. 

The corresponding eigenvectors are \cite{Gamayun,Naumis2020},

\begin{equation}
     |\Psi_{\alpha}^{\alpha'}(\vec{k})\rangle= |\Psi_{\alpha}(\vec{k})\rangle \otimes |\Psi_{\alpha'}(\vec{k})\rangle
\end{equation}
where $|\Psi_{\alpha}\rangle$ is a single-valley eigenvector for pristine graphene. More explicitly, defining $\theta=\tan^{-1}k_y /k_x$, the eigenvectors can be written in terms of the cone and band indexes as, 
\begin{equation}
    |\Psi^{\beta}_{\alpha}(\vec{k})\rangle=\frac{1}{2}\left( \begin{array}{cccc}
       1,  & \alpha e^{i \theta_\vec{k}}, & \alpha\beta e^{i \theta_\vec{k}}, & \beta e^{2 i \theta_\vec{k}}
    \end{array}\right)^T
\end{equation}
where the cone index is $\beta=\alpha\alpha'$.

\section{Dynamic polarizability of Kek-Y distorted graphene}\label{Sec.Polarizability}

We first study the dynamical polarizability up to lowest order in perturbation theory, defined by the bare bubble Feynman diagram, known as the Lindhard formula \cite{Fetter,Mahan}. The dynamical polarizability is a function of the wavevector magnitude $q$ and frequency $\omega$ and can be written as \cite{Sarma,Wunsch,Peres},

\begin{equation}\label{eq:DynamicalPi}
    \Pi(q,\omega)=-g_s\int\frac{d^2\vec{k}}{4\pi^2}\sum_{\substack{\alpha,\alpha',\\ \beta,\beta'}}\frac{f^{\beta}_{\vec{k}\alpha}-f^{\beta'}_{\vec{k}'\alpha' }}{\omega+ \epsilon_{\vec{k}\alpha}^\beta-\epsilon_{\vec{k}'\alpha'}^{\beta'}+i\eta}F_{\alpha\alpha'}^{\beta\beta'}(\vec{k},\vec{k}')
\end{equation}

where  $\eta$ is a small self-energy added for convergence, $g_s=2$ is the spin degeneracy, $\vec{k}'=|\vec{k}+\vec{q}|$,  $f^{\beta}_{\vec{k}\alpha}=1/[e^{\beta(\epsilon^\beta_{\vec{k}\alpha}-\mu)}+1]$ is the Fermi-Dirac distribution and
\begin{eqnarray}\label{eq:Fdef}
    F_{\alpha\alpha'}^{\beta\beta'}(\vec{k},\vec{k}')=\frac{1}{4}&&(1+\alpha\alpha'\cos\theta_{kk'} )  \nonumber\\
    &&\times(1+\alpha\alpha'\beta\beta'\cos\theta_{kk'})
\end{eqnarray}
being the form factor or scattering probability $|\langle\Psi_{\alpha'}^{\beta'}(\vec{k}+\vec{q})|\Psi_{\alpha}^{\beta}(\vec{k})\rangle|^2$ between states with momentum $\vec{k}$ and $\vec{k}'=\vec{k}+\vec{q}$.

For simplicity, we  consider the zero temperature case in which the Fermi-Dirac distribution becomes a step function. It is convenient to separate the polarizability in two components,
\begin{equation}
    \Pi(q,\omega)=\Pi^+(q,\omega)+\Pi^-(q,\omega),
\end{equation}
where $\Pi^+$ contains all terms with $f_{\vec{k}+}^\beta$ and $\Pi^-$ contains all terms with $f_{\vec{k}-}^\beta$. Then the full expressions for the components can be written as,
\begin{eqnarray}\label{Eq:Pi+}
\Pi^+(q,\omega)=-D_0\sum_{\beta\beta'\alpha}\int&&\frac{d^2\vec{k}}{4\pi}\bigg[\frac{F_+^{\beta\beta'}(\vec{k},\vec{k}')}{g_-^{\beta\beta'}+\alpha\omega_0^+}  \nonumber \\
&&+\frac{F_-^{\beta\beta'}(\vec{k},\vec{k}')}{g_+^{\beta\beta'}+\alpha\omega_0^+}\bigg]\Theta(k_\beta-k),
\end{eqnarray}

\begin{eqnarray}\label{Eq:Pi-}
\Pi^-(q,\omega)=D_0\sum_{\beta\beta'\alpha}\int\frac{d^2\vec{k}}{4\pi}\frac{F_-^{\beta\beta'}(\vec{k},\vec{k}')}{g_+^{\beta\beta'}+\alpha\omega_0^+}\Theta(\Lambda-k).
\end{eqnarray}
where $g_\pm^{\beta\beta'}=s_\beta k\pm s_{\beta'}k'$, $\omega_0=\omega/\mu$ and $\omega_0^+=\omega_0+i\eta$. $\Theta(x)$ denotes the Heaviside function, with $k_\beta=(1+\beta\Delta_0)^{-1}$ and  $\Lambda$ an arbitrary, high momentum cutoff. $D_0$ is the density of states of pristine graphene evaluated at the Fermi level $\mu=v_0 k_F$ \cite{KATSNELSON200720,Foa},
\begin{equation}
D_0=2k_F/\pi v_0.
\end{equation}
with $k_F$ being the Fermi momentum of nondistorted graphene. Notice also that momentum has been scaled by $k
_F$ so $k$ and $k'$ are unitless and $k'=|k+q/k_F|$. In the following we will use a tilde to denote the scaled polarizability  $\tilde{\Pi}(q,\omega)=\Pi(q,\omega)/D_0$.

\begin{figure}[t]
\includegraphics[width=0.45\textwidth]{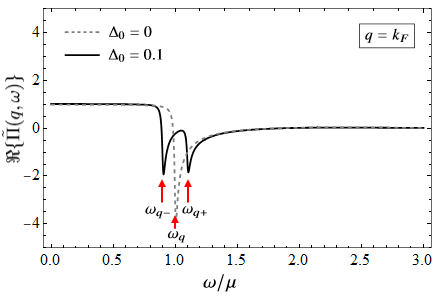}
\includegraphics[width=0.45\textwidth]{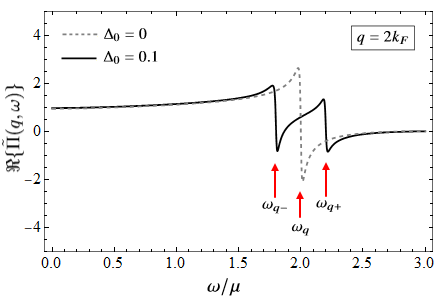}
\caption{\label{Fig:Re_Pi} Real part of the dynamical polarizability $\tilde{\Pi}(q,\omega)$  of graphene with ($\Delta_0=0.1$) and without ($\Delta_0=0$) kekul\'e distortion for $q=k_F$  (top)  and  $q=2k_F$ (bottom). The  jump in graphene at $\omega_q=v_0 q$, splits to frequencies $\omega_{q\pm}= v_0 (1\pm\Delta_0)q$ as a result of the kekul\'e distortion. These plots were obtained by numerical evaluation of Eqs. (\ref{Eq:Pi+}) and (\ref{Eq:Pi-}).}
\end{figure}

\begin{figure}[h]
\includegraphics[width=0.46\textwidth]{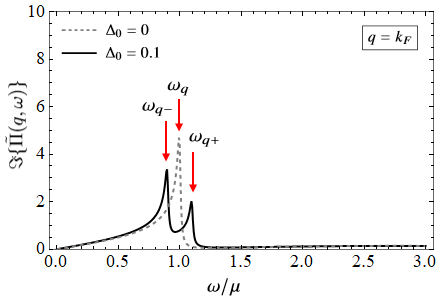}
\includegraphics[width=0.46\textwidth]{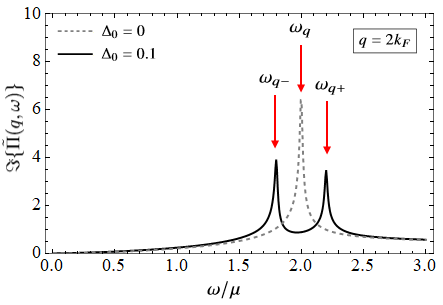}
\caption{\label{Fig:Im_Pi} Imaginary part of the dynamical polarizability $\tilde{\Pi}(q,\omega)$ of graphene with ($\Delta_0=0.1$) and without ($\Delta_0=0$) kekul\'e distortion  for  $q=k_F$ (top) and  $q=2k_F$ (bottom). The absorption peak in graphene at $\omega_q=v_0q$, splits to frequencies $\omega_{q\pm}= v_0 (1\pm\Delta_0)q$ as a result of the kekul\'e distortion. These plots were obtained by numerical evaluation of Eqs. (\ref{Eq:Pi+}) and (\ref{Eq:Pi-}).}
\end{figure}
 \par

Figs. \ref{Fig:Re_Pi} and \ref{Fig:Im_Pi} show the real and imaginary parts of $\tilde{\Pi}(q,\omega)$ for Kek-Y distorted graphene for different wavevectors. These plots were obtained by numerical evaluation of Eqs. (\ref{Eq:Pi+}) and (\ref{Eq:Pi-}). As a comparison, in Figs. \ref{Fig:Re_Pi} and \ref{Fig:Im_Pi}  we also include the results for pristine graphene. From these plots, it is clear that the main effect introduced by the kekul\'e distortion is the splitting of the response in two branches. While pristine graphene's spectrum exhibits a peak at $\omega_q=v_0q$, the kekul\'e distorted graphene exhibits two peaks at $\omega_{q\pm}=v_0(1\pm\Delta_0)q$, therefore making evident the presence two species of massless Dirac fermions.

Since the spectrum resembles that of pristine graphene after being split at two frequencies, it would be worthwhile to understand whether this is just  the sum of the polarizabilities of graphene for each valley shifted in frequency from one another by some frequency proportional to $\Delta_0$.
This seems plausible since, as can be confirmed from the definition of the polarizability in Eq. (\ref{eq:DynamicalPi}), a change in the Fermi velocity $v_0\rightarrow  v_0(1\pm\Delta_0)$ is equivalent to a shift in frequency $\omega\rightarrow \omega/(1\pm\Delta_0)$ and an overall scaling of $1/(1\pm\Delta_0)$. To  answer this question, we rewrite Eq. (\ref{eq:DynamicalPi}) taking advantage of the known polarizability of pristine graphene. First we notice that the scattering probability for kekul\'e patterned graphene, $F_{\alpha\alpha'}^{\beta\beta'}(\Vec{k},\Vec{k}')$, can be written in terms of graphene's single-valley scattering probability $F_{\alpha\alpha'}^g(\Vec{k},\Vec{k}')=\frac{1}{2}(1+\alpha\alpha'\cos\theta_{kk'})$ \cite{Sarma,Wunsch,Peres} as,
\begin{equation}\label{eq:SeparatedF}
    F_{\alpha\alpha'}^{\beta\beta'}(\vec{k},\vec{k}')=\delta_{\beta,\beta'}F_{\alpha\alpha'}^g(\vec{k},\vec{k}')-\beta\beta'\bigg(\frac{q\sin\varphi}{2|\vec{k}+\vec{q}|}\bigg)^2,
\end{equation}
where $\varphi$ is the angle between $\Vec{k}$ and $\Vec{q}$. Using Eqs. (\ref{eq:DynamicalPi}) and (\ref{eq:SeparatedF}), it can be verified that in the limit of $\Delta_0\rightarrow0$, the polarizability contains the factor $\sum_{\beta\beta'} F_{\alpha\alpha'}^{\beta\beta'}=2F_{\alpha\alpha'}^g$, that is, in the limit of nondistorted graphene the valley degeneracy is restored and the scattering function reduces to two times (one-per valley) the single-valley scattering function $F_{\alpha\alpha'}^g$, recovering the known expression for the polarizability of pristine graphene \cite{Sarma,Wunsch,Peres}, as expected. However, when the kekul\'e distortion is introduced, the full scattering function $F_{\alpha\alpha'}^{\beta\beta'}$ contains an additional term [see Eq. (\ref{eq:SeparatedF})], which leads to a new component related to intervalley transport in the polarization.
To see this, we rewrite the full polarizability of Kek-Y distorted graphene in terms of that of pristine graphene by plugging Eq. (\ref{eq:SeparatedF}) into Eqs. (\ref{Eq:Pi+}) and (\ref{Eq:Pi-}). After this, the the polarizability can be written as,

\begin{figure*}[t]
\includegraphics[width=1\textwidth]{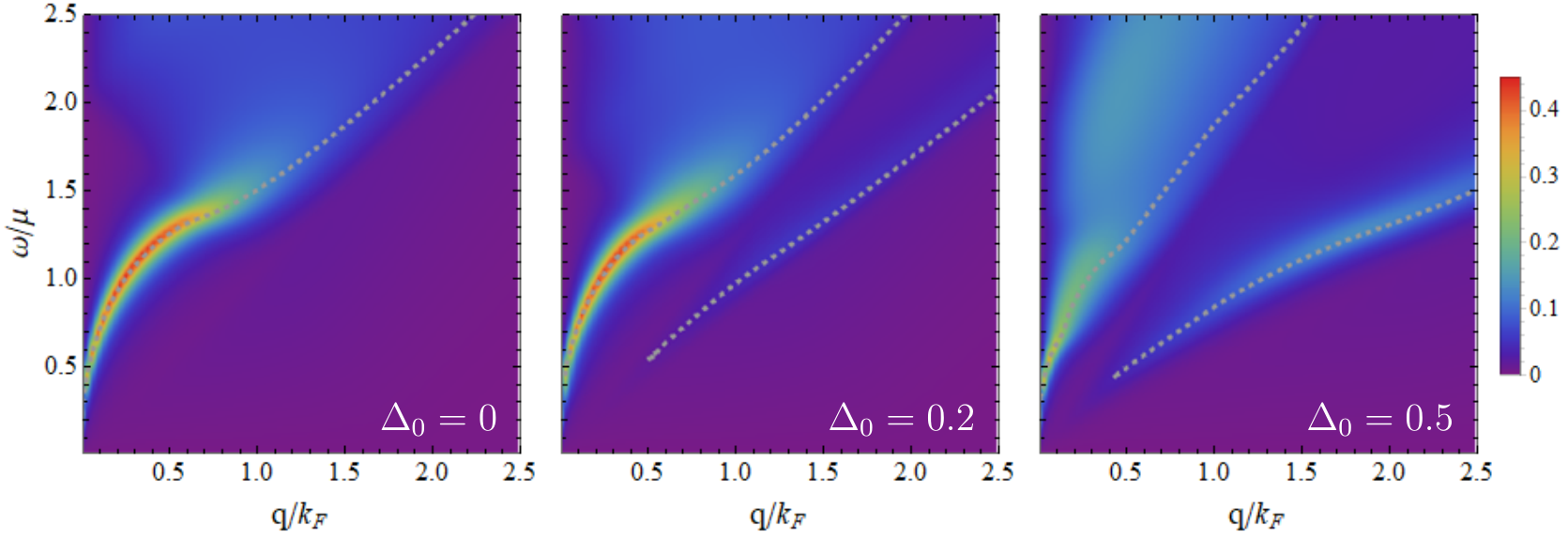}
\caption{\label{Fig:Plasmons} Plasmon dispersion obtained from the full polarizability  (dashed lines) and the loss function $\mathcal{L}=-\Im\{1/\epsilon(q,\omega)\}$ (color plot) within RPA for multiple values of the coupling amplitude. For a nonzero coupling amplitude
a second branch is introduced in the Landau damping region. However, at small values of the coupling amplitude ($\Delta_0=0.2$) the plasmon dispersion in the stable domain is not affected significantly. At a high value of $\Delta_0=0.5$ the stability of the main branch is reduced substantially. Plots are scaled as $\log_{100}[1+\mathcal{L}]$.}
\end{figure*}

\begin{eqnarray}\label{eq:SimplifiedPi}
    \tilde{\Pi}(q,\omega)= \frac{\tilde{\Pi}^{g}(q,\omega_+)}{2s_-}+&&\frac{\tilde{\Pi}^{g}(q,\omega_-)}{2s_+}\nonumber \\&&\quad\quad+\Delta_0\tilde{\Pi}^Y(q,\omega).
\end{eqnarray}
where we used the definition,
\begin{equation}
    \omega_{+}=\frac{\omega}{s_{-}},  \qquad    \omega_{-}=\frac{\omega}{s_{+}}\
\end{equation}

or, up to first order $\Delta_0$, $\omega_\pm=(1\pm\Delta_0)\omega$, so the relation $\omega_+\geqslant\omega\geqslant\omega_-$ is attained. Here $\Pi^{g}(q,\omega)$ is the well known full dynamical polarizability of graphene \cite{Wunsch,Peres,Sarma}, while $\tilde{\Pi}^Y(q,\omega)$ is an intervalley component introduced by the kekul\'e distortion. The first two terms correspond to the polarizability of each cone, which is the same as that for graphene except for the appropriate change in Fermi velocities $v_0\rightarrow v_0(1\pm\Delta_0)$, which is equivalent to a change in frequency $\omega\rightarrow \omega_\mp$ and an overall scaling of $1/s_\pm$. Therefore, in the limit of $\Delta_0\rightarrow0$, this two terms coincide and add up to the well known polarizability of graphene \cite{Wunsch,Sarma,Peres}, while the intervalley term $\Delta_0\tilde{\Pi}^Y(q,\omega)$ vanishes, resulting in $\tilde{\Pi}(q,\omega)=\tilde{\Pi}^g(q,\omega)
$.
Apart from explicitly showing the breaking of the valley degeneracy introduced by the kekul\'e distortion, substantial insight can be gained from Eq. (\ref{eq:SimplifiedPi}). 
First, since $\tilde{\Pi}^g(q,\omega)$ exhibits a peak at $\omega=v_0q$,  $\tilde{\Pi}^g(q,\omega_\pm)$ must instead exhibit peaks at $\omega_\pm=v_0q$, that is, $\omega=v_0(1\pm\Delta_0)q$, which is indeed confirmed in Figs. \ref{Fig:Re_Pi} and \ref{Fig:Im_Pi}. Second, the polarizability for kekul\'e distorted graphene is not merely given by adding the polarizabilities per cone with a relative frequency shift between them, as might appear at first sight from Figs. \ref{Fig:Re_Pi} and \ref{Fig:Im_Pi}. Indeed, there is an additional intervalley component, given by 
\begin{eqnarray}\label{Eq.Pi_Y}
\tilde{\Pi}^Y(q,\omega)=&&\sum_{\alpha,\beta}\beta\int\frac{d^2\vec{k}}{8\pi}\frac{q^2\sin^2\varphi}{k'} \nonumber \\ && \times\bigg[\frac{\Theta(k_\beta-k)}{(v_\beta k-k'+\alpha\omega_0^+)^2+(\Delta_0 k')^2}\nonumber \\
&& +\frac{\Theta(\Lambda-k)-\Theta(k_\beta-k)}{(v_\beta k+k'+\alpha\omega_0^+)^2+(\Delta_0 k')^2}\bigg].
\end{eqnarray}

It should be emphasized that this term does not appear in pristine graphene, as transitions from one cone to the other are canceled out in the low-energy approximation. It can be seen that in  the limit $\Delta_0 \rightarrow 0$ this component vanishes in Eq. (\ref{eq:SimplifiedPi}). While the analytic expressions for $\Pi^g(q,\omega)$ are well known \cite{Wunsch,Peres,Sarma}, the solution for $\Pi^Y(q,\omega)$ is quite complicated and here we only report numerical solutions for it. However, an analytical solution in the limit of $q\rightarrow0$ can be obtained from the expressions of the local conductivity previously reported in Ref. \cite{Naumis2020}. Nevertheless, as seen in Figs. \ref{Fig:Re_Pi} and \ref{Fig:Im_Pi}, at finite $q$ and $\omega$ the contribution of $\Pi^Y(q,\omega)$ in the real and imaginary parts of $\Pi(q,\omega)$ is a small perturbation, while the main effect of the kekul\'e distortion is displayed by a ``split" of graphene's spectrum, as accounted by the rescaled terms with $\Pi^g(q,\omega_+)$ and $\Pi^g(q,\omega_-)$; while graphene's spectrum exhibits jumps at $\omega_q=v_0 q$,  kekul\'e-distorted graphene will exhibit such jumps at $\omega_{q\pm}=v_0(1\pm\Delta_0)q$.

As we discuss in the following sections, however,  while the intervalley component $\tilde{\Pi}^Y$ is negligible at finite $q$ and $\omega$, it becomes apparent in the limit of $\omega\rightarrow0$ through the static screening (see Sec. \ref{Sec.Screening})  and in the limit of $q\rightarrow0$ through the local optical conductivity (see Sec. \ref{Sec.Conductivity}). We extend this discussion in the following sections.

\section{\label{Sec.Plasmons}Plasmons}
Accounting for the Coulomb interaction makes possible the study of collective charge excitations, known as plasmons \cite{Grigorenko2012,Ju2011,Peres,deAbajo,Bao,Tame2013,Jablan}.
Within RPA, the plasmon dispersion $\omega_p$ is given by the roots of the dielectric function $\epsilon(q,\omega)$ obtained from the self-consistent RPA polarization \cite{Fetter,Mahan},
\begin{eqnarray}
\Pi_{\text{RPA}}(q,\omega)=-\frac{\Pi(q,\omega)}{1+v_q\Pi(q,\omega)}
\end{eqnarray}
as,
\begin{eqnarray}
\epsilon(q,\omega)=1+v_q\Pi(q,\omega),
\end{eqnarray}
where $v_q=e^2/2\kappa_0 q$ \cite{Wunsch,Sarma,Peres}. Notice that a negative sign has been introduced to make $\Pi_{\text{RPA}}$ coincide with the definitions on \cite{Wunsch,Peres}. Additionally, the dispersion and damping of graphene plasmons can be uncovered from the loss function \cite{Peres,Agarwal},
\begin{eqnarray}
\mathcal{L}=-\Im\{1/\epsilon(q,\omega)\},
\end{eqnarray}
which takes maximum values where there is high probability of energy loss due to the excitation of stable plasmonic modes, and falls to zero as it enters the interband Landau damping domain, where plasmons become unstable \cite{Peres}.
In Fig. \ref{Fig:Plasmons} we plot and compare the loss function for  nondistorted graphene ($\Delta_0=0$) and for  kekul\'e-patterned graphene ($\Delta_0>0$). In order to focus solely on the effect of $\Delta_0$ we use $\alpha\equiv e^2/4\pi \kappa_0 v_0=2.5$ and $\epsilon_0=1$ \cite{Wunsch,Sarma} in all calculations. We also superimpose the dispersion curves obtained from the roots of $\epsilon(q,\omega)\approx 1+v_q \Re\{\Pi(q,\omega)\}$ (assuming weak damping \cite{Wunsch,Peres}).
It can be seen that at a coupling amplitude of  $\Delta_0=0.2$, the general low-$q$ plasmon dispersion of graphene in the stability region is practically not affected by the kekul\'e distortion, while a second branch of the plasmonic dispersion appears  in $\omega_{q-}<\omega<\omega_{q+}$. However, the system exhibits optical absorption in this domain, since $\Im\{\Pi(q,\omega)\}\neq0$ (see Fig. \ref{Fig:Im_Pi}) and therefore, plasmons in this region of the $\omega-q$ space are not stable, decaying quickly into electron-hole excitations. We further notice that this second branch appears for $q/k_F\gtrsim 0.5$, and therefore it must be related to nonlocal effects \cite{Peres}.
The fact that a small $\Delta_0$ has no effect in the low-$q$ plasmonic dispersion is in agreement with recent calculations of the Kubo conductivity for Kek-Y distorted graphene \cite{Naumis2020}, where it was found that a small $\Delta_0$ has no effect in the Drude peak, which determines the optical response and the plasmonic dispersion in the $q\rightarrow0$ approximation \cite{Peres}. Additionally, the loss function presents two lines at $\omega=v_0(1\pm\Delta_0)q$ instead of a single one at $\omega=v_0q$, resembling the energy dispersion of  kekul\'e-distorted graphene, as expected. 
At a large value of $\Delta_0=0.5$,  it can be seen that the stability of the main branch has been reduced substantially. This is due to the fact that the optical gap is reduced by the coupling amplitude $\Delta_0$ (see Sec. \ref{Sec.Conductivity}).

\section{\label{Sec.Screening}Static screening}
The static response is obtained from the dynamical polarizability in the $\omega \rightarrow 0$ limit. Here we denote each component by $\Pi^\pm(q,\omega\rightarrow0)=\Pi^\pm(q)$. The real components can be obtained from Eqs. (\ref{Eq:Pi+}) and (\ref{Eq:Pi-}) as
\begin{eqnarray}\label{Eq:RePi+}
\Re\{\tilde{\Pi}^+(q)\}=-&&\sum_{\beta\beta'\alpha}\int\frac{d^2\vec{k}}{2\pi} \frac{(v_\beta k -\alpha v_{\beta'}k')}{(v_\beta k -\alpha v_{\beta'}k')^2 +\eta^2} \nonumber
\\ && \times F_\alpha^{\beta\beta'}(\vec{k},\vec{k}')\Theta(k_\beta-k), 
\end{eqnarray}
\begin{eqnarray}\label{Eq:RePi-}
\Re\{\tilde{\Pi}^-(q)\}=&&\sum_{\beta\beta'}\int\frac{d^2\vec{k}}{2\pi} \frac{(v_\beta k +v_{\beta'}k')}{(v_\beta k + v_{\beta'}k')^2 +\eta^2}
 \nonumber \\ &&\times F_-^{\beta\beta'}(\vec{k},\vec{k}')\Theta(\Lambda-k).
\end{eqnarray}
It is easy to see from Eqs. (\ref{Eq:Pi+}) and (\ref{Eq:Pi-}) that the imaginary parts are zero (as expected in the $\omega\rightarrow 0$ limit), therefore, we have that $\Pi^+(q)=\Re\{\Pi^+(q)\}$ and
$\Pi^-(q)=\Re\{\Pi^-(q)\}$.

\begin{figure}
\hspace*{-1cm}
\includegraphics[width=0.4\textwidth]{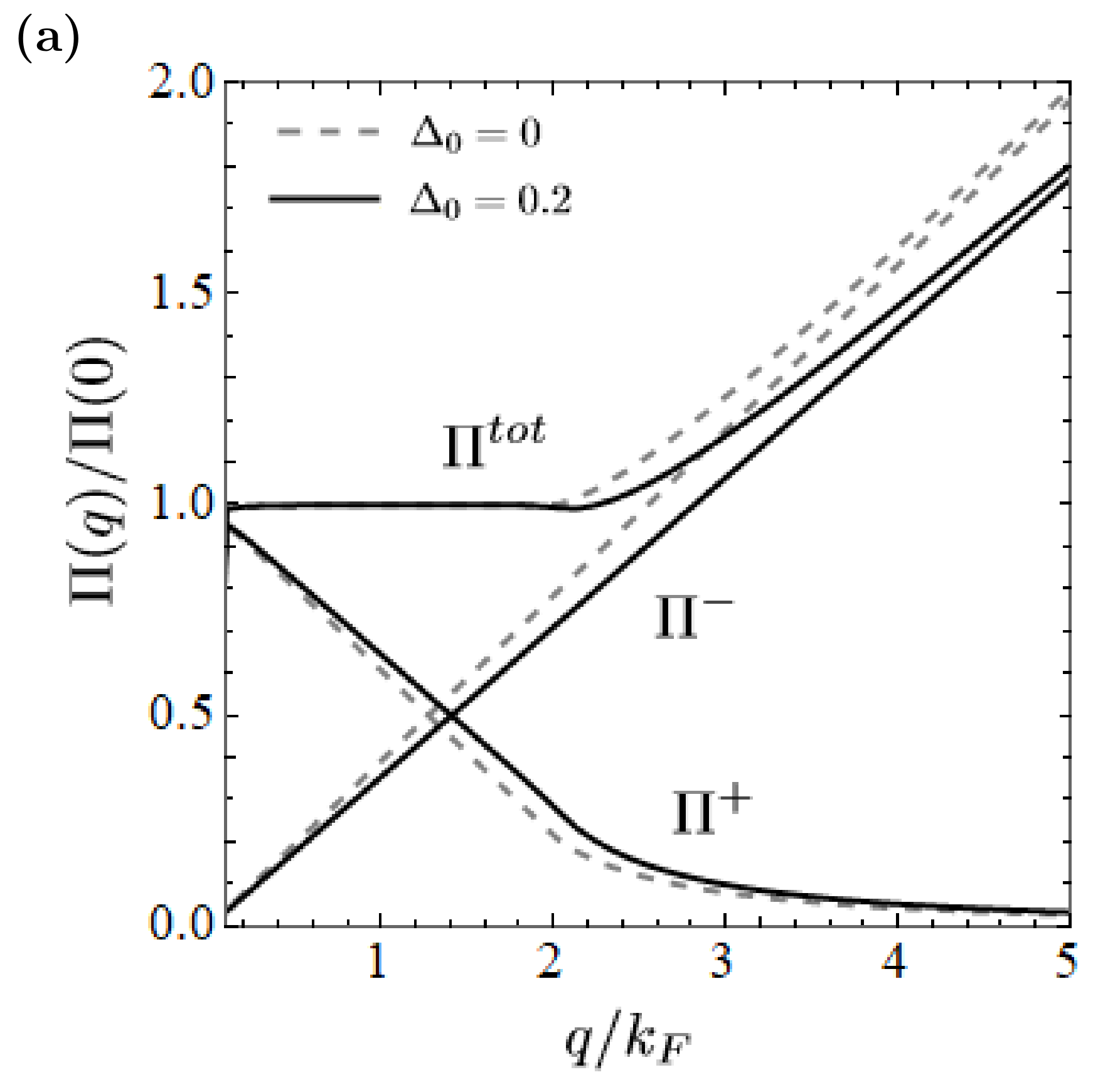}
\hspace*{-1cm}
\includegraphics[width=0.4\textwidth]{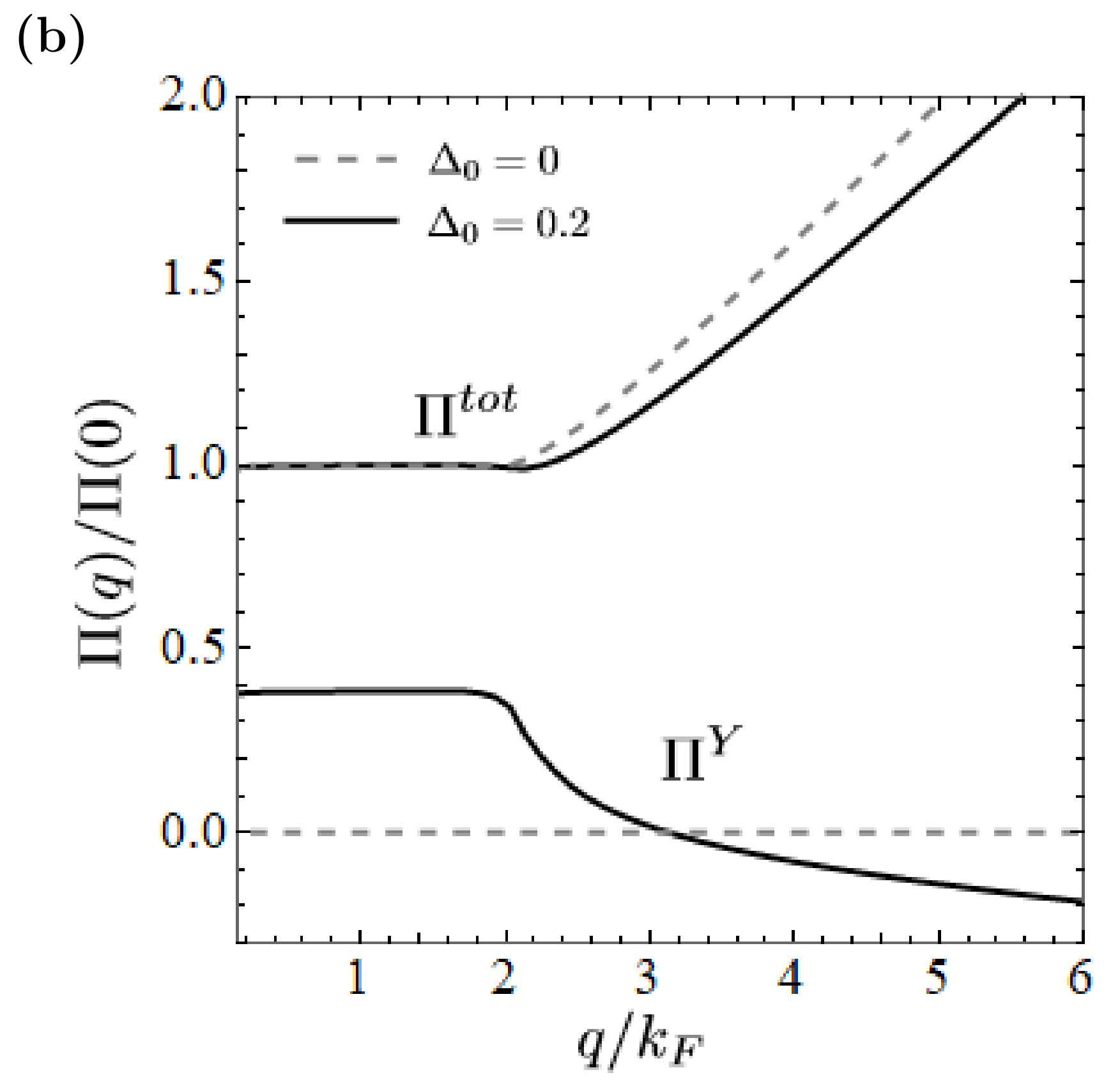}
\caption{\label{Screening} The total static polarizability $\Pi(\omega \rightarrow 0,q)$ in non-distorted ($\Delta_0=0$) and kekul\'e-distorted graphene ($\Delta_0=0.2$) also compared to (a) the doped $\Pi^+$ and undoped $\Pi^-$ components, and (b) the intervalley component $\Pi^Y$ [see Eq. (\ref{eq:SimplifiedPi})]. The curves were obtained from a numerical evaluation of Eqs. (\ref{Eq:RePi+}) and (\ref{Eq:RePi-}).
}
\end{figure}

The total static screening at $q=0$ coincides with the density of states at $\mu$,
\begin{equation}\label{Eq:Density_of_states}
    \Pi(0)=\frac{\mu}{\pi v_0^2s_+^2}+\frac{\mu}{\pi v_0^2  s_-^2}\approx (1+3\Delta_0^2)D_0
\end{equation}
which in the case of $\Delta_0\rightarrow0$ reduces to the density of states of non-distorted graphene. 
In Fig. \ref{Screening} we plot and compare $\Pi(q)$ for both the nondistorted ($\Delta_0=0$) and distorted ($\Delta_0>0$) graphene. We show  the doped $\Pi^+(q)$ and undoped $\Pi^-(q)$ components (Fig. \ref{Screening}a), as well as  the intervalley component $\Pi^Y(q)$ (Fig. \ref{Screening}b). For $q<2k_F$, in pristine graphene, as in the normal 2D electron gas (described by a quadratic energy dispersion), the static polarizability is equal to the density of states at the Fermi level, $\Pi(0)$ \cite{Sarma,AndoReview}. It can be seen in Fig. \ref{Screening} that for $\Delta_0>0$ this is still the case. Notice however, that since the density of states increases with $\Delta_0$ [see Eq. (\ref{Eq:Density_of_states})], the (nonscaled) static polarizability in Kek-Y distorted graphene actually takes a higher value. On the other hand, for $q>2k_F$, while the static polarizability of the normal 2D electron gas falls off from $\Pi(0)$ to zero \cite{AndoReview}, that of graphene increases linearly with $q$ \cite{Sarma}. Since this characteristic behavior is a consequence of the gapless dispersion of graphene, which is preserved in the Kek-Y distorted phase, the same linear dependence should be observed. This is confirmed to be the case in Fig. \ref{Screening}. We notice, however, that for $\Delta_0>0$ the (nonscaled) static polarizability exhibits a larger slope, increasing the effective dielectric constant at short wavelengths, which implies a suppression of the effective interaction in Kek-Y distorted graphene \cite{Sarma}. In Fig. \ref{Screening}b we also compare the total polarizability to the intervalley component $\Pi^Y$. It is interesting to notice that $\Pi^Y(q)$ shows resemblance to the static polarizability of a normal 2D material, taking a constant value for $q<2k_F$ and falling rapidly for $q>2k_F$, although this component takes negative values at large wavevectors, instead of falling to zero. It should be noticed, however, that this component enters the total polarizability as $\Delta_0\Pi^Y(q)$ [see Eq. (\ref{eq:SimplifiedPi})], therefore being a small contribution to the full static response.

\section{Optical conductivity}\label{Sec.Conductivity}

\begin{figure}[h]
\hspace*{-0.65cm}
\includegraphics[width=0.47\textwidth]{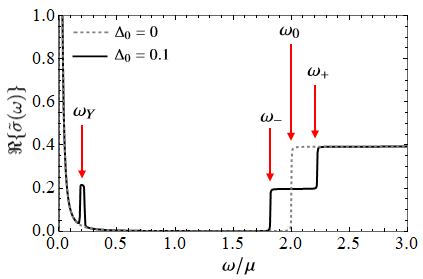}
\hspace*{-.2cm}
\includegraphics[width=0.45\textwidth]{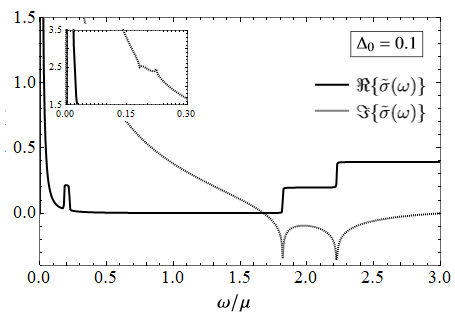}
\caption{\label{Fig:Sigma}Top: Real part of the local-optical conductivity $\tilde{\sigma}(\omega)$ 
 for nondistorted ($\Delta_0=0$) and Kek-Y distorted ($\Delta_0=0.1$) graphene. The kekul\'e distortion splits the step-like interband conductivity of graphene into two steps as a consequence of the broken valley degeneracy, and a resonance peak at $\omega_Y=(\omega_+-\omega_-)/2$ arises due to intervalley absorption. Bottom: The real and imaginary parts of $\tilde{\sigma}(\omega)$ for Kek-Y distorted graphene. The inset shows the singularities of $\Im\{\tilde{\sigma}\}$ due to the peak at $\omega_Y$.}
\end{figure}

In this section we present a short discussion on the signatures of the kekul\'e distortion in the optical conductivity of graphene. The (local) optical conductivity can be obtained from the polarizability \cite{Peres} as
\begin{eqnarray}
\tilde{\sigma}(\omega)=\lim_{q\rightarrow0}i\frac{ -\omega/\mu}{(q/k_F)^2}\tilde{\Pi}(q,\omega),
\end{eqnarray}
which we have plotted and compared to that of nondistorted graphene in Fig. \ref{Fig:Sigma} (to obtain the conductivity in conventional units, it suffices to multiply $\tilde{\sigma}$ by a factor of $4e^2/h$).

As recent calculations for the local conductivity using the Kubo formula have shown \cite{Naumis2020}, a tunable absorption peak due to intervalley transitions is exhibited at a frequency $ \omega_Y\approx2\mu\Delta_0$. We find that, indeed, this peak is introduced by the intervalley component $\Pi^Y(q,\omega)$ in Eq. (\ref{Eq.Pi_Y}). Furthermore, graphene's characteristic step-like absorption spectrum (starting at $\omega_0=2\mu$) splits into two half-steps (starting at $\omega_\pm=2\mu$), as was first noticed in the context of the Landauer formalism applied to Kek-Y distorted graphene nanoribbons \cite{Elias2020}. Additionally, this two-step optical conductivity (and the energy dispersion) of Kek-Y distorted graphene holds resemblance to that of a  $3/2$-pseudospin Dirac semimetal \cite{Dora2011}, which might indicate that modulation in the lattice could even change the effective pseudospin of the system. Also, a very similar two-step absorption was recently shown to be introduced by electron-hole asymmetry in one of the 2D phases of boron \cite{Verma}, which have been drawing a lot of interest due to their remarkable anisotropic transport properties \cite{Feng,Jafari2020,Lherbier_2016,Lozovik}.  We notice that this split of the conductivity into two half-steps is made evident in our expression for the polarizability in Eq. (\ref{eq:SimplifiedPi}), which also allows us to obtain the respective characteristic frequencies. Since graphene's polarizability $\tilde{\Pi}^g(0,\omega)$ exhibits (through the conductivity) the step at $\omega=2\mu$, the terms in the polarizability of Kek-Y distorted graphene, $\tilde{\Pi}^g(0,\omega_\pm)$, must exhibit the step at $\omega_\pm=2\mu$, that is $\omega=2\mu(1\pm\Delta_0)$. This is in agreement with the curves in Fig. \ref{Fig:Sigma}.

For a further discussion on the intervalley and intravalley transport in Kek-Y distorted graphene we refer the reader to Ref. \cite{Naumis2020}, where analytical expressions for $\tilde{\sigma}(\omega)$ can also be found.
Here we will focus the rest of our discussion on the fact that, up to first order in $\Delta_0$, 
\begin{equation}\label{Eq:Beating_freq}
    \omega_Y=\frac{\omega_+-\omega_-}{2},
\end{equation}
which is reminiscent of the general beating effect that results from the interference of two waves with close but different frequencies $f_1$ and $f_2$, leading to the modulation of the resulting wave by an envelope of frequency $f=(f_2-f_1)/2$.  In acoustics and optics, the difference between two slightly different frequencies is known as the ``beat frequency'', since the  modulation of the resulting wave can be perceived as beats or pulses \cite{AcousticsBook,Patorski:11}.  This phenomenon is specially relevant in the context of space-modulated 2D materials, like twisted bilayer graphene, where moir\'e beating patterns take place when there is a slight mismatch between the periodicities of the two lattices \cite{Koshino2019,Ochoa2020,Cao2018}, leading to a larger-scale (lower-frequency) spatial modulation of the system and introducing several novel physical properties \cite{Cao2018,Naik,Wu_2020}. 
Eq. (\ref{Eq:Beating_freq}) indicates that a related effect is present in the optical conductivity of Kek-Y modulated graphene; the breaking of the valley degeneracy due to the keku\'e distortion introduces two close but different frequencies  ($\omega_+$ and $\omega_-$) at which the onset of absorption in each valley occurs (see Fig. \ref{Fig:Sigma}) and the resonant frequency at which intervalley absorption takes place is given by the corresponding ``beat frequency'' in Eq. (\ref{Eq:Beating_freq}). In this way, the beating effect in the system originating from the presence of two slightly different scales (here defined by $\omega_+$ and $\omega_-$) is manifested through the optical conductivity by introducing a resonance peak at the beat frequency, which corresponds to the intervalley absorption.  

\section{Conclusion}
Using the  RPA approximation (Lindhard formula), we calculated the dynamic and static polarizability of kekul\'e distorted graphene and investigated the signatures of the broken valley degeneracy through the dielectric response, the plasmonic dispersion, the static screening and the optical conductivity.  As a consequence of the valley-dependent Fermi velocity, the dielectric spectrum splits, making evident the presence of two species of massless Dirac fermions.
Furthermore, the kekul\'e modulation introduces a second branch to the plasmonic dispersion of graphene. We used the loss function to study the plasmon stability at each frequency and wavevector. In the static limit, it was found that the static screening is increased at small wavelengths, implying a suppression of the effective interaction. We also discussed the optical conductivity, in which the characteristic step-like spectrum of graphene  splits into two half steps due to the onset of absorption in each valley, occurring at different characteristic frequencies. This 
effect is akin to that observed in a 3/2-pseudospin Dirac semimetal \cite{Dora2011} suggesting a possible change in the effective pseudospin.

Lastly, we described an absorption phenomenon where a resonance peak related to intervalley transport emerges at a beat frequency determined by the characteristic frequencies of each valley.
 We expect some of these signatures to be present in other space-modulated 2D materials, as strained graphene \cite{NaumisReview,Taboada2017,Naumis_Mapping}, twisted-angle graphene \cite{MOGERA2020470,Wu_2020}, patterned graphene nanoribbons \cite{NaumisTerrones2009}  or even in modulated quasicrystals \cite{Naumis_Thorpe}. Our work suggests that simple optical or electrical measurements can be suitable to detect this kind of modulation in 2D materials.\\
 \par

We thank DGAPA-PAPIIT project  IN102620. S. A. H. acknowledges financial support from CONACyT.

\bibliography{References_Dielectric}
\end{document}